\input amstex
\documentstyle{amsppt}
%
\catcode`@=11
\redefine\output@{%
  \def\break{\penalty-\@M}\let\par\endgraf
  \ifodd\pageno\global\hoffset=105pt\else\global\hoffset=8pt\fi  
  \shipout\vbox{%
    \ifplain@
      \let\makeheadline\relax \let\makefootline\relax
    \else
      \iffirstpage@ \global\firstpage@false
        \let\rightheadline\frheadline
        \let\leftheadline\flheadline
      \else
        \ifrunheads@ 
        \else \let\makeheadline\relax
        \fi
      \fi
    \fi
    \makeheadline \pagebody \makefootline}%
  \advancepageno \ifnum\outputpenalty>-\@MM\else\dosupereject\fi
}
\catcode`@=\active
\nopagenumbers
\def\negskp{\hskip -2pt}

\def\const{\operatorname{const}}

\def\divr{\operatorname{div}}
\def\blue#1{#1}
\catcode`#=11\def\diez{#}\catcode`#=6
\catcode`_=11\def\podcherkivanie{_}\catcode`_=8
\catcode`~=11\catcode`~=\active
\def\mycite#1{\cite{\blue{#1}}\immediate\special{ps:
     ShrHPSdict begin /ShrBORDERthickness 0 def}}
\def\myciterange#1#2{\cite{\blue{#2}}\immediate\special{ps:
     ShrHPSdict begin /ShrBORDERthickness 0 def}}
\def\mytag#1{%
    \tag#1}
\def\mythetag#1{\thetag{\blue{#1}}\immediate\special{ps:
     ShrHPSdict begin /ShrBORDERthickness 0 def}}
\def\myrefno#1{\no#1}
\def\myhref#1#2{\blue{#2}\immediate\special{ps:
     ShrHPSdict begin /ShrBORDERthickness 0 def}}
\def\myEarXivlink{\myhref{http://arXiv.org}{http:/\negskp/arXiv.org}}
\def\myGeoCities{\myhref{http://www.geocities.com}{GeoCities}}
\def\mytheorem#1{\csname proclaim\endcsname{Теорема #1}}
\def\mytheorem#1{\csname proclaim\endcsname{Theorem #1}}
\def\mythetheorem#1{\blue{#1}\immediate\special{ps:
     ShrHPSdict begin /ShrBORDERthickness 0 def}}
\font\eightcyr=wncyr8
\pagewidth{360pt}
\pageheight{606pt}
\topmatter
\title
On the geometry of a dislocated medium.
\endtitle
\author
Jeffrey Comer, Ruslan Sharipov
\endauthor
\address Physics Department, The University of Akron, Akron, OH 
\endaddress
\email 
\myhref{mailto:jeffcomer\@gmail.com}{jeffcomer\@gmail.com}
\endemail
\address Department of Mathematics, Bashkir State University,
Frunze street 32, 450074 Ufa, Russia
\endaddress
\email \vtop to 30pt{\hsize=280pt\noindent
\myhref{mailto:R\podcherkivanie Sharipov\@ic.bashedu.ru}
{R\_\hskip 1pt Sharipov\@ic.bashedu.ru}\newline
\myhref{mailto:r-sharipov\@mail.ru}
{r-sharipov\@mail.ru}\newline
\myhref{mailto:ra\podcherkivanie sharipov\@lycos.com}{ra\_\hskip 1pt
sharipov\@lycos.com}\vss}
\endemail
\urladdr
\myhref{http://www.geocities.com/r-sharipov}
{http:/\negskp/www.geocities.com/r-sharipov}
\endurladdr
\abstract
    Purely real space versions of the differential equations describing
the kinematics of a dislocated crystalline medium are considered. The
differential geometric structures associated with them are revealed.
\endabstract
\subjclassyear{2000}
\subjclass 74A05, 74C15, 74C20, 74D10\endsubjclass
\endtopmatter
\loadbold
\TagsOnRight
\document

\head
1. Introduction.
\endhead
    In \mycite{1} a phenomenological approach to the nonlinear theory
of plasticity in glasses, pitches, and soft polymers was suggested 
(see also \mycite{2}). Currently, we have no direct microscopic support 
for the results of \mycite{1} since microscopic mechanisms of plasticity
in amorphous materials are not yet completely understood, especially if 
one needs an exact quantitative description (see papers \mycite{3} and
\mycite{4} where some approaches are developed, but this is by no means
the ultimate theory). Having no direct way, one should maneuver
choosing a detour, a roundabout course to the goal. For the theory of
plasticity this course goes through the theory of dislocations (see
\myciterange{5}{5--7}). The matter is that dislocations provide a
microscopic mechanism explaining the plasticity of crystals. Relying
on the integrity of the nature, one can expect that the plasticity
phenomenon in crystals and in amorphous materials are described 
similarly.\par
    The paper \mycite{5} is a review of the basics. There the nonlinear
elastic and plastic deformation tensors $\hat\bold G$ and $\check\bold G$
for a crystalline medium are defined, and the following differential equations for them are derived: 
$$
\gather
\hskip -2em
\gathered
\frac{\partial\hat G_{kq}}{\partial t}+\sum^3_{r=1}v^r\,
\nabla_{\!r}\hat G_{kq}=-\sum^3_{r=1}\nabla_{\!k}v^r\,
\hat G_{rq}-\sum^3_{r=1}\hat G_{kr}\,\nabla_{\!q}v^r+\\
+\sum^3_{r=1}\theta^{\,r}_k\,\hat G_{rq}+\sum^3_{r=1}
\hat G_{kr}\,\theta^{\,r}_q,
\endgathered
\mytag{1.1}\\
\vspace{2ex}
\hskip -2em
\frac{\partial\check G^{\,k}_i}{\partial t}+
\sum^3_{r=1}v^r\,\nabla_{\!r}\check G^{\,k}_i=
\sum^3_{r=1}\left(\check G^{\,r}_i\,\nabla_{\!r}v^k
-\nabla_{\!i}v^r\,\check G^{\,k}_r\right)-\sum^3_{r=1}
\theta^{\,k}_r\,\check G^{\,r}_i.
\mytag{1.2}
\endgather
$$
Here $v^1,\,v^2,\,v^3$ are the components of the velocity vector $\bold v$
of a point of the medium. By $\nabla$ in \mythetag{1.1} and \mythetag{1.2}
we denote the covariant differentiation. In Cartesian coordinates $x^1,\,
x^2,\,x^3$ the covariant derivative $\nabla_{\!i}$ coincides with
$\partial/\partial x^i$. However, for the sake of generality below we
use curvilinear coordinates $y^1,\,y^2,\,y^3$. In this case $\nabla_{\!i}$
is written through the Christoffel symbols (see \mycite{8}) of the standard
Euclidean metric in the space $\Bbb E$ (the real space, where the dynamics
of any medium occurs):
$$
\hskip -2em
\Gamma^k_{ij}=\sum^3_{r=1}\frac{g^{kr}}{2}\!\left(\frac{\partial g_{rj}}
{\partial y^i}+\frac{\partial g_{ir}}{\partial y^j}-\frac{\partial g_{ij}}
{\partial y^k}\right)\!.
\mytag{1.3}
$$\par
    It is remarkable that the same differential equations \mythetag{1.1} 
and \mythetag{1.2} govern the evolution of $\hat\bold G$ and $\check\bold G$ in amorphous media (see \mycite{1} and \mycite{2}). The difference is that
the tensorial parameter $\boldsymbol\theta$ for amorphous materials is
introduced empirically (see \mycite{1}), while for crystalline materials we
have the formula
$$
\hskip -2em
\theta^{\,r}_q=-\sum^3_{i=1}\hat S^r_i\,j^{\,i}_q+\sum^3_{i=1}
\sum^3_{p=1}v^p\,\hat S^r_i\,(\nabla_{\!p}\hat T^i_q-\nabla_{\!q}
\hat T^i_p)
\mytag{1.4}
$$
expressing $\boldsymbol\theta$ through other tensorial parameters 
of a medium: $\hat\bold T$ and $\bold j$ (see \mycite{6} and \mycite{7} 
for more details); $\hat\bold T$ is called the {\it incompatible distorsion
tensor\/} and $\bold j$ is the tensor of the {\it density of the Burgers
vector flow}. Unlike $\boldsymbol\theta$, both $\hat\bold T$ and 
$\bold j$ are double space tensorial quantities: their upper index $i$
in \mythetag{1.4} is associated with the Burgers space. The tensor field
$\hat\bold S$ in \mythetag{1.4} is expressed through $\hat\bold T$ as
the inverse matrix: $\hat\bold S=\hat\bold T^{-1}$. Its lower index
$i$ is associated with the Burgers space.\par
    The Burgers space $\Bbb B$ is introduced in \mycite{5} as a container
for Burgers vectors. It is a copy of the real space $\Bbb E$ filled with the
infinite defect-free crystalline grid of that material which we have in the
real space. The copy of the Euclidean metric $\bold g$ in the Burgers space
is denoted by $\overset\sssize\star\to{\bold g}$. The Burgers space $\Bbb B$
is usually equipped with Cartesian coordinates $x^1,\,x^2,\,x^3$. Therefore,
we have
$$
\xalignat 2
&\overset\sssize\star\to g\vphantom{g}_{ij}=\const,
&&\overset\sssize\star\to g\vphantom{g}^{ij}=\const.
\endxalignat
$$
Suppose that $\bold X$ is a double space tensor field in $\Bbb E$, and 
assume that $i$ and $j$ are its indices associated with the Burgers space:
$$
\hskip -2em
X^{\ldots\ i\,\ldots\ldots\ldots}_{\ldots\ldots\ldots\ j\,\ldots}
=X^{\ldots\ i\,\ldots\ldots\ldots}_{\ldots\ldots\ldots\ j\,\ldots}
(y^1,y^2,y^3).
\mytag{1.5}
$$
Applying $\hat\bold S$ and $\hat\bold T$ to the components \mythetag{1.5}
of the tensor field $\bold X$, one can convert $i$ and $j$ into the real
space indices $p$ and $q$ respectively:
$$
\align
&\hskip -2em
X^{\ldots\,p\ \ldots\ldots\ldots}_{\ldots\ldots\ldots\,j\,\ldots}
=\sum^3_{i=1}X^{\ldots\,i\,\ldots\ldots\ldots}_{\ldots\ldots\ldots
\,j\,\ldots}\ \hat S^p_i,
\mytag{1.6}\\
&\hskip -2em
X^{\ldots\,i\,\ldots\ldots\ldots}_{\ldots\ldots\ldots\,q\,\ldots}
=\sum^3_{j=1}X^{\ldots\,i\,\ldots\ldots\ldots}_{\ldots\ldots\ldots
\,j\,\ldots}\ \hat T^j_q.
\mytag{1.7}
\endalign
$$
The elastic deformation tensor $\hat\bold G$ is produced from
$\overset\sssize\star\to{\bold g}$ according to the recipe
\mythetag{1.7}:
$$
\hskip -2em
\hat G_{pq}=\sum^3_{i=1}\sum^3_{j=1}\overset\sssize\star\to
g\vphantom{g}_{ij}\ \hat T^i_p\ \hat T^j_q.
\mytag{1.8}
$$
The purely real space tensors $\bold J$ and $\hat\bold Z$ are produced
according to the recipe \mythetag{1.6}:
$$
\xalignat 2
&\hskip -2em
J^p_q=\sum^3_{i=1}\hat S^p_i\ j^{\,i}_{\,q},
&&\hat Z^{\,r}_{pq}=\sum^3_{i=1}\hat S^r_i\ \nabla_{\!p}\hat T^i_q.
\mytag{1.9}
\endxalignat
$$
The purely real space tensors $\bold R$ and $\hat\bold R$ are defined
similarly (see \mycite{7}):
$$
\xalignat 2
&\hskip -2em
R^p_q=\sum^3_{i=1}\hat S^p_i\ \rho^{\,i}_{\,q},
&&\hat R^{\,r}_{pq}=\sum^3_{i=1}\hat S^r_i\,(\nabla_{\!p}
\hat T^i_q-\nabla_{\!q}\hat T^i_p).
\mytag{1.10}
\endxalignat
$$
Here $\rho^{\,i}_{\,q}$ are the components of the dual space tensor field
$\boldsymbol\rho$, this tensor field is defined as the {\it density of
Burgers vector\/} for the dislocations in a crystal (see \mycite{5}). 
The formulas \mythetag{1.9} and \mythetag{1.10} express our intension
to write a substantial part of the theory in terms of purely real space
tensor fields. This goal was declared in \mycite{7}. The other goal of the
present paper is to reveal geometric structures hidden underneath the
theory of dislocations.\par
\head
2. The elastic deformation metric\\
and associated connection with torsion.
\endhead
    Let's consider the elastic deformation tensor $\hat\bold G$. It is
defined by formula \mythetag{1.8}. Since $\det\hat\bold T\neq 0$, this
formula determines a Riemannian metric in $\Bbb E$ other than the basic
Euclidean metric $\bold g$. Let's call it the {\it elastic deformation
metric\/} and denote it by $\hat\bold G$. Then remember the following
differential equation derived in \mycite{7}:
$$
\hskip -2em
\nabla_{\!p}\,\hat G_{qk}=\sum^3_{r=1}\hat Z^{\,r}_{pq}\,\hat G_{rk}
+\sum^3_{r=1}\hat Z^{\,r}_{pk}\,\hat G_{qr}.
\mytag{2.1}
$$
Expressing $\nabla_{\!p}$ in \mythetag{2.1} through the partial
derivative $\partial/\partial y^p$, we get
$$
\hskip -2em
\frac{\partial\hat G_{qk}}{\partial y^p}-\sum^3_{r=1}\Gamma^r_{pq}\,
\hat G_{rk}-\sum^3_{r=1}\Gamma^r_{pk}\,\hat G_{qr}=\sum^3_{r=1}
\hat Z^{\,r}_{pq}\,\hat G_{rk}+\sum^3_{r=1}\hat Z^{\,r}_{pk}\,
\hat G_{qr}.
\mytag{2.2}
$$
Here the Christoffel symbols $\Gamma^r_{pq}$ and $\Gamma^r_{pk}$ are
determined by the formula \mythetag{1.3}. Comparing the left and right
hand sides of \mythetag{2.2}, we see that it is convenient to introduce
the other set of Christoffel symbols:
$$
\hskip -2em
\hat\Gamma^k_{ij}=\Gamma^k_{ij}+\hat Z^k_{ij}.
\mytag{2.3}
$$
The Christoffel symbols \mythetag{2.3} define the other connection in 
the real space $\Bbb E$ (different from the standard Euclidean connection
\mythetag{1.3}) and the other covariant differentiation $\hat\nabla$. In
terms of $\hat\nabla$ the equation \mythetag{2.1} is written as
$$
\hskip -2em
\hat\nabla_{\!p}\,\hat G_{qk}=0.
\mytag{2.4}
$$
The equations like \mythetag{2.4} are known as concordance conditions
for metrics and connections. For instance, $\bold g$ and $\nabla$ are
concordant and \mythetag{1.3} is derived from the concordance condition
$\nabla_{\!p}\,g_{qk}=0$ (see \mycite{9}).\par
    Unlike $\Gamma^k_{ij}$, the newly introduced Christoffel symbols
\mythetag{2.3} are not symmetric, i\.\,e\. $\hat\Gamma^k_{ij}\neq
\hat\Gamma^k_{j\,i}$. Therefore, they define a nonzero torsion:
$$
\hskip -2em
\hat T^k_{ij}=\hat\Gamma^k_{ij}-\hat\Gamma^k_{j\,i}.
\mytag{2.5}
$$
For the {\it torsion tensor\/} with the components \mythetag{2.5} we 
use the same symbol $\hat\bold T$ as for the distorsion tensor 
above\footnote{\ The matter is that in geometry the symbol {\eightcyr\char '074}T{\eightcyr\char '076} is a typical notation for a torsion.}. However,
one should remember that they are two different things: the distorsion
tensor $\hat\bold T$ is a double space tensor with two indices, while 
the torsion tensor $\hat\bold T$ is a purely real space tensor with three
indices. Substituting \mythetag{2.3} into \mythetag{2.5}, for the torsion
tensor $\hat\bold T$ we derive:
\adjustfootnotemark{-1}
$$
\hskip -2em
\hat T^k_{ij}=\hat Z^k_{ij}-\hat Z^k_{j\,i}=\hat R^k_{ij}.
\mytag{2.6}
$$
Now let's remember that the tensor fields $\bold R$ and $\hat\bold R$
with the components \mythetag{1.10} are related to each other by the
following equality derived in \mycite{7}:
$$
\hskip -2em
\hat R^{\,k}_{ij}=\sum^3_{s=1}\sum^3_{r=1}\omega_{sij}\ g^{sr}\,R^k_r.
\mytag{2.7}
$$
Here $\omega_{sij}$ are the components of the {\it volume tensor\/}
$\boldsymbol\omega$ (see \mycite{8} and \mycite{9} for more details).
Substituting \mythetag{2.7} into \mythetag{2.6}, we obtain the equality
$$
\hskip -2em
\hat T^{\,k}_{ij}=\sum^3_{s=1}\sum^3_{r=1}\omega_{sij}\ g^{sr}\,R^k_r.
\mytag{2.8}
$$
\mytheorem{2.1} The torsion tensor of the connection \mythetag{2.3}
is determined by the density of Burgers vector of a dislocated medium 
through the equality \mythetag{2.8}.
\endproclaim
    Apart from the torsion tensor, any connection possesses another 
tensorial parameter --- the {\it curvature tensor}. It is given by
the formula
$$
\hskip -2em
\hat R^p_{q\,ij}=\frac{\partial\hat\Gamma^p_{\!j\,q}}
{\partial y^i}-\frac{\partial\hat\Gamma^p_{\!i\,q}}
{\partial y^j}+\sum^3_{m=1}\hat\Gamma^m_{\!j\,q}\,
\hat\Gamma^p_{\!i\,m}-\sum^3_{m=1}\hat\Gamma^m_{\!i\,q}
\,\hat\Gamma^p_{\!j\,m}.
\mytag{2.9}
$$
The standard Euclidean connection \mythetag{1.3} is a flat connection,
this means that its curvature tensor is identically equal to zero:
$$
\hskip -2em
R^p_{q\,ij}=\frac{\partial\Gamma^p_{\!j\,q}}
{\partial y^i}-\frac{\partial\Gamma^p_{\!i\,q}}
{\partial y^j}+\sum^3_{m=1}\Gamma^m_{\!j\,q}\,
\Gamma^p_{\!i\,m}-\sum^3_{m=1}\Gamma^m_{\!i\,q}
\,\Gamma^p_{\!j\,m}=0.
\mytag{2.10}
$$
Substituting \mythetag{2.3} into \mythetag{2.9} and taking into account 
\mythetag{2.10}, we derive
$$
\hskip -2em
\hat R^p_{q\,ij}=\nabla_{i}\hat Z^p_{j\,q}-\nabla_{\!j}\hat Z^p_{i\,q}
+\sum^3_{m=1}\hat Z^m_{j\,q}\ \hat Z^p_{i\,m}-\sum^3_{m=1}\hat Z^m_{i\,q}
\ \hat Z^p_{j\,m}.
\mytag{2.11}
$$
In order to calculate $\nabla_{\!i}\hat Z^p_{j\,q}$ and $\nabla_{\!j}\hat
Z^p_{i\,q}$ in \mythetag{2.11} we use the second formula \mythetag{1.9}:
$$
\nabla_{\hskip -1pt i}\hat Z^p_{j\,q}=\sum^3_{m=1}\nabla_{\hskip -1pt
i}(\hat S^p_m\ \nabla_{\!j}\hat T^m_q)=\sum^3_{m=1}\hat S^p_m\ 
\nabla_{\hskip -1pt i}\nabla_{\!j}\hat T^m_q+\sum^3_{m=1}
\nabla_{\hskip -1pt i}\hat S^p_m\ \nabla_{\!j}\hat T^m_q.
\quad
\mytag{2.12}
$$
The matrix $\hat\bold S$ is inverse to $\hat\bold T$. Therefore, we have
the equality
$$
\hskip -2em
\nabla_{\hskip -1pt i}\hat S^p_m=-\sum^3_{n=1}\sum^3_{k=1}\hat S^p_n
\ \nabla_{\hskip -1pt i}\hat T^n_k\ \hat S^k_m.
\mytag{2.13}
$$
Now from \mythetag{2.12} and \mythetag{2.13} for $\nabla_{\!i}\hat
Z^p_{j\,q}$ we derive
$$
\hskip -2em
\nabla_{\hskip -1pt i}\hat Z^p_{j\,q}=\sum^3_{m=1}\hat S^p_m\ 
\nabla_{\hskip -1pt i}\nabla_{\!j}\hat T^m_q-\sum^3_{m=1}
\hat Z^p_{i\,m}\,\hat Z^m_{j\,q},
\mytag{2.14}
$$
and then for $\nabla_{\!j}\hat Z^p_{i\,q}$ we write by analogy
$$
\hskip -2em
\nabla_{\!j}\hat Z^p_{i\,q}=\sum^3_{m=1}\hat S^p_m\ 
\nabla_{\!j}\nabla_{\hskip -1pt i}\hat T^m_q-\sum^3_{m=1}
\hat Z^p_{j\,m}\,\hat Z^m_{i\,q}.
\mytag{2.15}
$$
Due to \mythetag{2.14} and \mythetag{2.15} the above expression 
\mythetag{2.11} is transformed to
$$
\hat R^p_{q\,ij}=\sum^3_{m=1}\hat S^p_m\,(\nabla_{\hskip -1pt i}\nabla_{\!j}\hat T^m_q-\nabla_{\!j}\nabla_{\hskip -1pt i}
\hat T^m_q).
$$
Due to the symmetry $\Gamma^k_{ij}=\Gamma^k_{\!j\,i}$ and due to
the flatness equality \mythetag{2.10} the covariant derivatives
$\nabla_{\hskip -1pt i}$ and $\nabla_{\!j}$ are commutative:
$\nabla_{\hskip -1pt i}\nabla_{\!j}=\nabla_{\!j}\nabla_{\hskip -1pt
i}$. Hence, we have
$$
\hskip -2em
\hat R^p_{q\,ij}=0.
\mytag{2.16}
$$
\mytheorem{2.2} Any dislocated crystalline medium is described
by a Riemannian metric $\hat\bold G$ (the elastic deformation metric) 
and by a non-symmetric flat connection $\hat{\boldsymbol\Gamma}$ being concordant with the metric $\hat\bold G$.
\endproclaim
    The proof is obvious. Indeed, the formula \mythetag{1.8} provides
a metric and \mythetag{2.3} provides a connection. Due to \mythetag{2.7}
this connection is non-symmetric. Due to \mythetag{2.4} this connection
is concordant with the metric \mythetag{1.8} and due to \mythetag{2.16}
it is flat, i\.\,e\. its curvature tensor is zero.\par
    The result of the theorem~\mythetheorem{2.2} is not new. As reported 
in \mycite{10}, Kondo, Bilby, Bullough, and Smith (see \mycite{11} and
\mycite{12}) in 1950s recognized that dislocations should be described in
terms of the differential geometry. However, their results are not widely
known to physicists and engineers. 
\head 
3. Reconstructing the distorsion.
\endhead
     According to the strategy declared in \mycite{7}, we are going
to replace double space tensors by purely real space \pagebreak 
tensorial parameters of a medium and write the complete set of the
differential equations in terms of these parameters. Suppose for a while
that this work is done and suppose that the elastic deformation metric
$\hat\bold G$ and the density of the Burgers vector in its real space form
$\bold R$ (see \mythetag{1.10}) are evaluated for some particular medium 
in some particular case. Then the torsion tensor $\hat\bold T$ is also 
known (see formulas \mythetag{2.7} and \mythetag{2.8}). The following
theorem says that the connection $\hat{\boldsymbol\Gamma}$ can be derived
from $\hat\bold G$ and $\bold R$.
\mytheorem{3.1} For any Riemannian metric $\hat\bold G$ and for any
tensorial field $\hat\bold T$ of the type $(1,3)$ there exists a unique
connection $\hat{\boldsymbol\Gamma}$ concordant with this metric and having
$\hat\bold T$ as its torsion tensor.
\endproclaim
     This theorem is a well-known geometric result. In symmetric case, 
i\.\,e\. if $\hat\bold T$ is zero, the connection $\hat{\boldsymbol\Gamma}$ is called the {\it standard metric connection} or the {\it Levi-Civita
connection} of the metric $\hat\bold G$ (see \mycite{9}).
\demo{Proof} The theorem \mythetheorem{3.1} is proved by deriving the
explicit formula for the components of $\hat{\boldsymbol\Gamma}$. Let's
denote by $\widetilde{\boldsymbol\Gamma}$ the symmetric part of 
$\hat{\boldsymbol\Gamma}$ and denote by $\widetilde{\bold G}$ the inverse 
matrix for the matrix of the metric tensor $\hat\bold G$:
$$
\xalignat2
&\hskip -2em\widetilde\Gamma^k_{ij}=\frac{\hat\Gamma^k_{ij}
+\hat\Gamma^k_{ji}}{2},
&&\widetilde G^{ij}=[\hat G^{-1}]^{ij}.
\mytag{3.1}
\endxalignat
$$
Traditionally, in differential geometry $\widetilde{\bold G}$ is called
the {\it inverse\/} or the {\it dual} metric tensor for $\hat\bold G$
and is denoted by the same symbol $\hat\bold G$. Here we cannot use 
such notations since, apart from $\hat\bold G$, we have the basic Euclidean
metric $\bold g$, hence, $\hat G^{ij}$ are implicitly determined by the
standard index raising procedure:
$$
\hat G^{ij}=\sum^3_{i=1}\sum^3_{j=1}g^{ip}\,g^{jq}\,\hat G_{pq}
\neq\widetilde G^{ij}.
$$
From \mythetag{2.5} and from the first equality \mythetag{3.1} we 
derive
$$
\hskip -2em
\hat\Gamma^k_{ij}=\widetilde\Gamma^k_{ij}+\frac{1}{2}\,\hat T^k_{ij}.
\mytag{3.2}
$$
Now let's write the concordance condition \mythetag{2.4} explicitly
using \mythetag{3.2}:
$$
\hskip -2em
\frac{\partial\hat G_{qk}}{\partial y^p}
-\sum^3_{r=1}\widetilde\Gamma^r_{pq}\,\hat G_{rk}
-\sum^3_{r=1}\widetilde\Gamma^r_{pk}\,\hat G_{qr}
=\frac{1}{2}\sum^3_{r=1}\hat T^r_{pq}\,\hat G_{rk}
+\frac{1}{2}\sum^3_{r=1}\hat T^r_{pk}\,\hat G_{qr}.
\mytag{3.3}
$$
Looking at \mythetag{3.3}, we see that it is convenient to denote
$$
\xalignat 2
&\hskip -2em
\widetilde\Gamma_{kij}=\sum^3_{r=1}\widetilde\Gamma^r_{ij}\,\hat G_{rk},
&&\widetilde T_{kij}=\sum^3_{r=1}\hat T^r_{ij}\,\hat G_{rk}.
\mytag{3.4}
\endxalignat
$$
Since $\hat\bold G$ is a non-degenerate matrix, the transformations 
\mythetag{3.4} are invertible:
$$
\xalignat 2
&\hskip -2em
\widetilde\Gamma^k_{pq}=\sum^3_{r=1}\widetilde\Gamma_{rpq}
\,\widetilde G^{rk},
&&\hat T^k_{pq}=\sum^3_{r=1}\widetilde T_{rpq}\,\widetilde G^{rk}.
\mytag{3.5}
\endxalignat
$$
Applying \mythetag{3.4} to \mythetag{3.3}, we can rewrite \mythetag{3.3}
in the following form:
$$
\hskip -2em
\widetilde\Gamma_{kpq}+\widetilde\Gamma_{qpk}=\frac{\partial
\hat G_{qk}}{\partial y^p}-\frac{1}{2}\,\widetilde T_{kpq}
-\frac{1}{2}\,\widetilde T_{qpk}.
\mytag{3.6}
$$
Performing two cyclic transpositions of indices $p\to q\to k\to p$ in \mythetag{3.6}, we produce the other two equalities from the equality
\mythetag{3.6}:
$$
\align
&\hskip -2em
\widetilde\Gamma_{pqk}+\widetilde\Gamma_{kqp}=\frac{\partial
\hat G_{kp}}{\partial y^q}-\frac{1}{2}\,\widetilde T_{pqk}
-\frac{1}{2}\,\widetilde T_{kqp},
\mytag{3.7}\\
&\hskip -2em
\widetilde\Gamma_{qkp}+\widetilde\Gamma_{pkq}=\frac{\partial
\hat G_{pq}}{\partial y^k}-\frac{1}{2}\,\widetilde T_{qkp}
-\frac{1}{2}\,\widetilde T_{pkq}.
\mytag{3.8}
\endalign
$$
Now let's add \mythetag{3.6} and \mythetag{3.7}, then subtract 
\mythetag{3.8} from the sum taking into account the symmetry 
of $\widetilde{\boldsymbol\Gamma}$ and the skew-symmetry of
$\widetilde{\bold T}$:
$$
2\,\widetilde\Gamma_{kpq}=\left(\frac{\partial\hat G_{qk}}
{\partial y^p}+\frac{\partial\hat G_{kp}}{\partial y^q}
-\frac{\partial\hat G_{pq}}{\partial y^k}\!\right)
-\widetilde T_{pqk}-\widetilde T_{qpk}.
$$
From this equality, applying \mythetag{3.5}, we derive the following
explicit formula for $\widetilde\Gamma^k_{pq}$:
$$
\hskip -2em
\widetilde\Gamma^k_{pq}=\sum^3_{r=1}\frac{\widetilde G^{kr}}{2}
\!\left(\frac{\partial\hat G_{qr}}
{\partial y^p}+\frac{\partial\hat G_{rp}}{\partial y^q}
-\frac{\partial\hat G_{pq}}{\partial y^r}
-\widetilde T_{pqr}-\widetilde T_{qpr}\!\right).
\mytag{3.9}
$$
And finally, substituting \mythetag{3.9} into the formula \mythetag{3.2},
we get
$$
\hskip -2em
\aligned
\hat\Gamma^k_{ij}=&\sum^3_{r=1}\,\frac{\widetilde G^{kr}}{2}\!\left(
\frac{\partial\hat G_{jr}}{\partial y^i}+\frac{\partial\hat G_{ri}}
{\partial y^j}-\frac{\partial\hat G_{ij}}{\partial y^r}\!\right)-\\
&-\sum^3_{r=1}\sum^3_{s=1}\frac{\hat G_{is}\,\hat T^s_{jr}+\hat G_{js}
\,\hat T^s_{ir}}{2}\,\widetilde G^{kr}+\frac{1}{2}\,\hat T^k_{ij}\,.
\endaligned
\mytag{3.10}
$$
This is the explicit formula for the components of the connection
declared in the theorem~\mythetheorem{3.1}. Thus, its existence and
uniqueness is proved.\qed\enddemo
\subhead A remark\endsubhead Because of the formulas \mythetag{2.3}
and \mythetag{2.8} the equality \mythetag{3.10} is equivalent to the
equality \thetag{3.13} from \mycite{7}. Thus, the
theorem~\mythetheorem{3.1} yields a geometric interpretation for the
formula \thetag{3.13} derived in the previous paper \mycite{7}.\par
     Now let's return to the equations \mythetag{1.9}. Since $\hat
\bold S$ and $\hat\bold T$ are inverse to each other ($\hat\bold
S=\hat\bold T^{-1}$), the second equation \mythetag{1.9} is rewritten 
in the following form:
$$
\hskip -2em
\nabla_{\!p}\hat T^i_q=\sum^3_{r=1}\hat Z^{\,r}_{pq}\,\hat T^i_r.
\mytag{3.11}
$$
More explicitly this formula is written as
$$
\hskip -2em
\frac{\partial\hat T^i_q}{\partial y^p}-\sum^3_{r=1}\Gamma^r_{pq}\,
\hat T^i_r=\sum^3_{r=1}\hat Z^{\,r}_{pq}\,\hat T^i_r.
\mytag{3.12}
$$
Applying \mythetag{2.3} to \mythetag{3.12}, we can bring the equation
\mythetag{3.12} to the following one:
$$
\hskip -2em
\frac{\partial\hat T^i_q}{\partial y^p}=\sum^3_{r=1}\hat\Gamma^r_{pq}
\,\hat T^i_r.
\mytag{3.13}
$$
And finally, there is the most simple form of the equality \mythetag{3.13},
where $\hat\nabla_{\!p}$ is used:
$$
\hskip -2em
\hat\nabla_{\!p}\hat T^i_q=0.
\mytag{3.14}
$$
Note that the upper index $i$ does not affect the expansion of the
covariant derivatives $\nabla_{\!p}$ and $\hat\nabla_{\!p}$ in
\mythetag{3.11} and \mythetag{3.14}. This is because the distorsion 
tensor $\hat\bold T$ is a double space tensor and its upper index $i$ 
is associated with the Burgers space $\Bbb B$.\par
     The equations \mythetag{3.13} form the so-called {\it complete system 
of Pfaff equations}. The section~8 in 
Chapter~\uppercase\expandafter{\romannumeral 5} of the thesis \mycite{13}
and the Appendix A in \mycite{14} can be used as a brief introduction to the
theory of Pfaff systems. All complete Pfaff systems are overdetermined 
systems of partial differential equations. The compatibility equations form
the basic feature of Pfaff systems. In order to derive them for 
\mythetag{3.13} one should calculate the second order partial derivatives
using \mythetag{3.13}:
$$
\gather
\frac{\partial\hat T^i_r}{\partial y^p\,\partial y^q}=
\sum^3_{s=1}\frac{\partial\hat\Gamma^s_{qr}}{\partial y^p}\,\hat T^i_s
+\sum^3_{s=1}\hat\Gamma^s_{qr}\,\frac{\partial\hat T^i_s}{\partial y^p}
=\sum^3_{s=1}\frac{\partial\hat\Gamma^s_{qr}}{\partial y^p}\,\hat T^i_s
+\sum^3_{s=1}\sum^3_{m=1}\hat\Gamma^s_{qr}\,\hat\Gamma^m_{ps}\,
\hat T^i_m,\\
\vspace{2ex}
\hskip -2em
\frac{\partial\hat T^i_r}{\partial y^p\,\partial y^q}=\sum^3_{s=1}
\left(\frac{\partial\hat\Gamma^s_{qr}}{\partial y^p}+\sum^3_{m=1}
\hat\Gamma^m_{qr}\,\hat\Gamma^s_{pm}\right)\hat T^i_s.
\mytag{3.15}
\endgather
$$
Exchanging the indices $p$ and $q$ in \mythetag{3.15}, we derive
$$
\hskip -2em
\frac{\partial\hat T^i_r}{\partial y^q\,\partial y^p}=\sum^3_{s=1}
\left(\frac{\partial\hat\Gamma^s_{pr}}{\partial y^q}+\sum^3_{m=1}
\hat\Gamma^m_{pr}\,\hat\Gamma^s_{qm}\right)\hat T^i_s.
\mytag{3.16}
$$
Subtracting \mythetag{3.16} from \mythetag{3.15} and taking into account
\mythetag{2.9}, we get
$$
\hskip -2em
\sum^3_{s=1}\hat R^s_{rpq}\,\hat T^i_s=0.
\mytag{3.17}
$$
The equality \mythetag{3.17} should be fulfilled for any solution of the
Pfaff system \mythetag{3.13}. However, in our case we have the stronger
result given by the theorem~\mythetheorem{2.2} and the equality \mythetag{2.16}. Regarding the Pfaff equations the equality
\mythetag{2.16} is called the {\it compatibility condition} of the Pfaff
system.\par
     Let's fix some point $P_0$ within the continuous medium. Without loss
of generality we can assume that its curvilinear coordinates are equal to
zero: $y^i(P_0)=0$. Then we choose some constant matrix and denote its components by $\hat T^i_s(0)$. The equality
$$
\hat T^i_s(0,0,0)=\hat T^i_s(0).
$$
can be understood as the initial value condition for \pagebreak
the solution 
$\hat T^i_s(y^1,y^2,y^3)$ of the Pfaff equations \mythetag{3.13}. 
We prefer to write it in the following form:
$$
\hskip -2em
\hat T^i_s\,\vbox{\hrule width 0.5pt
height 8pt depth 8pt}_{\,\,P=P_0}=\hat T^i_s(0).
\mytag{3.18}
$$
\mytheorem{3.2} The initial value problem \mythetag{3.18} for the 
system of Pfaff equations \mythetag{3.13} has a unique local
solution\footnote{\ This means the solution in some neighborhood 
of the initial point $P_0$.} for any predefined matrix $\hat T^i_s(0)$ if 
and only if the compatibility condition \mythetag{2.16} is fulfilled.
\endproclaim
\adjustfootnotemark{-1}
     The theorem~\mythetheorem{3.2} is a standard fact of the theory of
Pfaff equations. We do not give the proof of this theorem here. The idea
for its proof can be found in the section~8 of the
Chapter~\uppercase\expandafter{\romannumeral 5} in \mycite{13}.\par
     In practice, we need only non-degenerate solutions of the Pfaff
equations \mythetag{3.13}, i\.\,e\. $\det\hat\bold T\neq 0$. Suppose 
we have two different solutions of the equations \mythetag{3.13}, we
denote them $\hat\bold T[1]$ and $\hat\bold T[2]$. Then their initial
values at the point $P_0$ are related to each other by means of some 
non-degenerate constant matrix $\bold O$:
$$
\hskip -2em
\hat T^i_s[2](0)=\sum^3_{j=1}\hat T^j_s[1](0)\ O^{\,i}_j.
\mytag{3.19}
$$
Since \mythetag{3.13} are linear equations and since the 
theorem~\mythetheorem{3.2} provides the uniqueness of the solution, 
the equality \mythetag{3.19} is fulfilled identically at all points:
$$
\hskip -2em
\hat T^i_s[2]=\sum^3_{j=1}\hat T^j_s[1]\ O^{\,i}_j.
\mytag{3.20}
$$
If both solutions $\hat\bold T[1]$ and $\hat\bold T[2]$ correspond to 
the same elastic deformation tensor $\hat\bold G$, then from \mythetag{1.8}
and \mythetag{3.20} we derive
$$
\hskip -2em
\overset\sssize\star\to g\vphantom{g}_{rs}=\sum^3_{i=1}\sum^3_{j=1}
\overset\sssize\star\to g\vphantom{g}_{ij}\ O^{\,i}_r\ O^{\,j}_s.
\mytag{3.21}
$$
This equality \mythetag{3.21} means that $\bold O$ is a constant orthogonal
matrix with respect to Euclidean metric in the Burgers space $\Bbb B$. Such
a matrix corresponds to a global rotation with or without reflection in
$\Bbb B$. Explaining the concept of the Burgers space in \cite{7}, we said
that it can be understood as an isometric copy of the real space $\Bbb E$.
From this point of view the global rotations and reflections are inessential
transformations in $\Bbb B$. Therefore, up to this inessential uncertainty
in defining Burgers vectors, now we have the one-to-one correspondence:
$$
\hskip -2em
\hat\bold T\rightleftarrows\hat\bold G,\,\bold R.
\mytag{3.22} 
$$
The double space tensorial field of distorsion $\hat\bold T$ defines the
pair of purely real space tensorial field: the elastic deformation tensor
$\hat\bold G$ and the tensor $\bold R$, which is the real space
representation of the Burgers vector density $\boldsymbol\rho$ (see formula
\mythetag{1.10} above). For the sake of brevity, from now on, we shall 
call $\bold R$ the {\it Burgers vector density} as well. The tensor
$\hat\bold G$ is derived from the distorsion $\hat\bold T$ by means of 
the formula \mythetag{1.8}. The formula for tensor $\bold R$ is more
complicated:
$$
\hskip -2em
R^{\,r}_k=\sum^3_{i=1}\sum^3_{s=1}\sum^3_{p=1}\sum^3_{q=1}\hat S^r_i\,
g_{sk}\ \omega^{spq}\ \nabla_{\!p}\hat T^i_q.
\mytag{3.23}
$$
The formulas \mythetag{1.8} and \mythetag{3.23} correspond to the upper
right arrow in \mythetag{3.22}. The lower left arrow in \mythetag{3.22}
goes through formula \mythetag{2.8}, through the theorem~\mythetheorem{3.1}
provided by the formula \mythetag{3.10}, and through solving the system
of Pfaff equations \mythetag{3.13}. As a result we recover the incompatible
distorsion tensor $\hat\bold T$ from $\hat\bold G$ and $\bold R$.\par
\subhead A remark\endsubhead Being produced from the single field
$\hat\bold T$, the tensor field $\hat\bold G$ and $\bold R$ are not 
absolutely independent. They are related to each other through the 
zero-curvature condition \mythetag{2.16}.\par
\head
4. Time evolution and consistence of the kinematic equations
in whole.
\endhead 
    The time evolution of the elastic deformation tensor $\hat\bold G$
is given by the equation \mythetag{1.1}. The time evolution for the tensor
$\bold R$ is given by the equation \thetag{4.11} from \mycite{7}:
$$
\gathered
\frac{\partial R^k_q}{\partial t}-\sum^3_{p=1}J^{\,k}_p\,R^p_q
-\sum^3_{m=1}\nabla_{\!m}v^k\,R^m_q-\sum^3_{m=1}\sum^3_{p=1}
v^p\,\hat Z^k_{mp}\,R^m_q\,+\\
+\sum^3_{m=1}\sum^3_{r=1}\sum^3_{s=1}g_{qm}\,\omega^{mrs}\,
\nabla_{\!r}J^{\,k}_{\,s}+
\sum^3_{m=1}\sum^3_{r=1}\sum^3_{s=1}\sum^3_{p=1}g_{qm}\,\omega^{mrs}\,
\hat Z^k_{rp}\,J^{\,p}_{\,s}=0.
\endgathered\quad
\mytag{4.1}
$$
Now let's remember the second relationship \mythetag{1.9} and apply it
to \mythetag{3.23}. Then
$$
\hskip -2em
R^p_q=\sum^3_{m=1}\sum^3_{r=1}\sum^3_{s=1}
g_{qm}\ \omega^{mrs}\ \hat Z^p_{rs}.
\mytag{4.2}
$$
Substituting \mythetag{4.2} into the second term of \mythetag{4.1} and
rearranging the terms, we get
$$
\hskip -2em
\gathered
\frac{\partial R^k_q}{\partial t}
-\sum^3_{m=1}\left(\nabla_{\!m}v^k+\sum^3_{p=1}\hat Z^k_{mp}
\,v^p\!\right)R^m_q\,+\\
+\sum^3_{m=1}\sum^3_{r=1}\sum^3_{s=1}g_{qm}\,\omega^{mrs}\!
\left(\nabla_{\!r}J^{\,k}_{\,s}+\sum^3_{p=1}\hat Z^k_{rp}\,J^{\,p}_{\,s}
-\hat Z^p_{rs}\,J^{\,k}_p\!\right)=0.
\endgathered
\mytag{4.3}
$$
Then we remember the relationship \mythetag{2.3} and replace $\nabla$
by $\hat\nabla$ in \mythetag{4.3}:
$$
\hskip -2em
\frac{\partial R^k_q}{\partial t}-\sum^3_{m=1}\hat\nabla_{\!m}v^k
\,R^m_q+\sum^3_{m=1}\sum^3_{r=1}\sum^3_{s=1}g_{qm}\,\omega^{mrs}
\,\hat\nabla_{\!r}J^{\,k}_{\,s}=0.
\mytag{4.4}
$$
In a similar way let's replace $\nabla$ by $\hat\nabla$ \pagebreak
in the evolution equation \mythetag{1.1} for the elastic deformation 
tensor $\hat\bold G$. Applying \mythetag{2.3} to \mythetag{1.1} and
using \mythetag{2.4}, we derive
$$
\gathered
\frac{\partial\hat G_{kq}}{\partial t}+\sum^3_{r=1}
\hat\nabla_{\!k}v^r\,\hat G_{rq}
+\sum^3_{r=1}\sum^3_{p=1}v^r\!\left(\hat Z^p_{rk}
-\hat Z^p_{k\,r}\right)\hat G_{pq}
+\sum^3_{r=1}\hat\nabla_{\!q}v^r\,\hat G_{kr}\,+\\
+\sum^3_{r=1}\sum^3_{p=1}v^r\!\left(\hat Z^p_{rq}
-\hat Z^p_{qr}\right)\hat G_{kp}
=\sum^3_{r=1}\theta^{\,r}_k\,\hat G_{rq}+\sum^3_{r=1}
\hat G_{kr}\,\theta^{\,r}_q.
\endgathered\quad
\mytag{4.5}
$$
Now let's compare \mythetag{1.9} with \mythetag{1.4}. As a result we write
\mythetag{1.4} as 
$$
\hskip -2em
\theta^{\,r}_q=-J^{\,r}_q+\sum^3_{p=1}v^p
\!\left(\hat Z^r_{p\,q}-\hat Z^r_{qp}\right).
\mytag{4.6}
$$
Applying \mythetag{4.6} to \mythetag{4.5}, we transform \mythetag{4.5} to
the following equation:
$$
\gathered
\frac{\partial\hat G_{kq}}{\partial t}+\sum^3_{r=1}
\hat\nabla_{\!k}v^r\,\hat G_{rq}
+\sum^3_{r=1}\hat\nabla_{\!q}v^r\,\hat G_{kr}\,+\\
+\sum^3_{r=1}J^{\,r}_k\,\hat G_{rq}+\sum^3_{r=1}
\hat G_{kr}\,J^{\,r}_q=0.
\endgathered\quad
\mytag{4.7}
$$\par
     The differential equations \mythetag{4.4} and \mythetag{4.7} describe
the time evolution of the tensor fields $\hat\bold G$ and $\bold R$. The
next step is to show that this time evolution is compatible with the
zero-curvature condition \mythetag{2.16}. The following commutation
relationships are derived from the definition of covariant derivatives by
direct calculations:
$$
\align
&\hskip -2em
[\partial_t,\hat\nabla_{\!p}]\,X^k=\sum^3_{s=1}
\frac{\partial\hat\Gamma^k_{ps}}{\partial t}\ X^s,
\mytag{4.8}
\\
&\hskip -2em
[\partial_t,\hat\nabla_{\!p}]\,X_k=-\sum^3_{s=1}
\frac{\partial\hat\Gamma^s_{pk}}
{\partial t}\ X_s.
\mytag{4.9}
\endalign
$$
Here $\partial_t=\partial/\partial t$, while $X^k$ and $X_k$ stand for the
components of arbitrary vectorial and covectorial fields respectively. In
order to calculate the time derivatives of the connection components in
\mythetag{4.8} and \mythetag{4.9} we use the equality \mythetag{2.4}
rewritten as
$$
\hskip -2em
\frac{\partial\hat G_{qk}}{\partial y^p}-\sum^3_{s=1}
\hat\Gamma^s_{p\,q}\,\hat G_{sk}-\sum^3_{s=1}
\hat\Gamma^s_{p\,k}\,\hat G_{qs}=0.
\mytag{4.10}
$$
For the sake of brevity in the further calculations we denote
$$
\xalignat 3
&\frac{\partial\hat\Gamma^k_{rs}}{\partial t}=\Dot{\Hat\Gamma}
\vphantom{\Gamma}^k_{rs},
&&\frac{\partial\hat G{rs}}{\partial t}=\Dot{\Hat G}\vphantom{G}_{rs},
&&\frac{\partial\hat T^k_{rs}}{\partial t}=\Dot{\Hat T}
\vphantom{T}^k_{rs},\quad
\endxalignat
$$
where $\hat T^k_{rs}$ are the components of the torsion tensor $\hat
\bold T$ related to $\bold R$ through \mythetag{2.8}. Now differentiating
\mythetag{4.10} with respect to $t$, we find
$$
\hskip -2em
\sum^3_{s=1}\Dot{\Hat\Gamma}\vphantom{\Gamma}^s_{p\,q}\,\hat G_{sk}
+\sum^3_{s=1}\Dot{\Hat\Gamma}\vphantom{\Gamma}^s_{p\,k}\,\hat G_{qs}
=\hat\nabla_{\!p}(\Dot{\Hat G}\vphantom{G}_{qk}).
\mytag{4.11}
$$
The equality \mythetag{4.11} is similar to \mythetag{3.3}. Therefore,
we shall treat it similarly. Let's apply \mythetag{3.1} and \mythetag{3.2}
to \mythetag{4.11}. As a result we get
$$
\hskip -2em
\gathered
\sum^3_{s=1}\Dot{\widetilde\Gamma}\vphantom{\Gamma}^s_{p\,q}\,\hat G_{sk}
+\sum^3_{s=1}\Dot{\widetilde\Gamma}\vphantom{\Gamma}^s_{p\,k}\,\hat G_{qs}=
\\
=\hat\nabla_{\!p}(\Dot{\Hat G}\vphantom{G}_{qk})-\frac{1}{2}\sum^3_{s=1}
\Dot{\Hat T}\vphantom{T}^s_{p\,q}\,\hat G_{sk}-\frac{1}{2}\sum^3_{s=1}
\Dot{\Hat T}\vphantom{T}^s_{p\,k}\,\hat G_{qs}.
\endgathered
\mytag{4.12}
$$
Now we introduce the following notations similar to \mythetag{3.4}:
$$
\xalignat 2
&\hskip -2em
\check\Gamma_{k\,ij}=\sum^3_{r=1}
\Dot{\widetilde\Gamma}\vphantom{\Gamma}^r_{ij}\,\hat G_{rk},
&&\check T_{k\,ij}=\sum^3_{r=1}\Dot{\Hat T}\vphantom{T}^r_{ij}
\,\hat G_{rk}.
\mytag{4.13}
\endxalignat
$$
Applying \mythetag{4.13} to \mythetag{4.12}, we strengthen the resemblance
of \mythetag{4.12} and \mythetag{3.3}:
$$
\hskip -2em
\check\Gamma_{kpq}+\check\Gamma_{qpk}
=\hat\nabla_{\!p}(\Dot{\Hat G}\vphantom{G}_{qk})-\frac{1}{2}
\,\check T_{kpq}-\frac{1}{2}\,\,\check T_{qpk}.
\mytag{4.14}
$$
Note that \mythetag{4.14} looks pretty like the equality \mythetag{3.6}.
Therefore, we can use the same arguments as in proving the 
theorem~\mythetheorem{3.1} and derive the following formula:
$$
\hskip -2em
\aligned
\Dot{\Hat\Gamma}\vphantom{\Gamma}^k_{ij}=&\sum^3_{r=1}\,\frac{\widetilde G^{kr}}{2}\!\left(\hat\nabla_{\!i}(\Dot{\Hat G}\vphantom{G}_{jr})
+\hat\nabla_{\!j}(\Dot{\Hat G}\vphantom{G}_{ri})
-\hat\nabla_{\!r}(\Dot{\Hat G}\vphantom{G}_{ij})\right)-\\
&-\sum^3_{r=1}\sum^3_{s=1}\frac{\hat G_{is}\,
\Dot{\Hat T}\vphantom{T}^s_{jr}+\hat G_{js}
\,\Dot{\Hat T}\vphantom{T}^s_{ir}}{2}\,\widetilde G^{kr}
+\frac{1}{2}\,\Dot{\Hat T}\vphantom{T}^k_{ij}\,.
\endaligned
\mytag{4.15}
$$
The covariant derivatives $\hat\nabla_{\!i}(\Dot{\Hat G}\vphantom{G}_{jr})$,
$\hat\nabla_{\!j}(\Dot{\Hat G}\vphantom{G}_{ri})$, and $\hat\nabla_{\!r}
(\Dot{\Hat G}\vphantom{G}_{ij})$ are given by the formula \mythetag{4.7}.
Applying this formula, we get
$$
\gathered
\hat\nabla_{\!i}(\Dot{\Hat G}\vphantom{G}_{jr})+
\hat\nabla_{\!j}(\Dot{\Hat G}\vphantom{G}_{ri})+
\hat\nabla_{\!r}(\Dot{\Hat G}\vphantom{G}_{ij})=
-\sum^3_{m=1}\left(\hat\nabla_{\!i}\hat\nabla_{\!j}\,v^m\,
+\right.\\
\left.+\,\hat\nabla_{\!j}\hat\nabla_{\!i}\,v^m
+\hat\nabla_{\!i}J^{\,m}_j+\hat\nabla_{\!j}J^{\,m}_i\right)
\hat G_{mr}+\sum^3_{m=1}\left(
[\hat\nabla_{\!r},\,\hat\nabla_{\!i}]v^m
+\hat\nabla_{\!r}J^{\,m}_i\,-\right.\\
\left.-\,\hat\nabla_{\!i}J^{\,m}_r\right)\hat G_{mj}
+\sum^3_{m=1}\left(
[\hat\nabla_{\!r},\,\hat\nabla_{\!j}]v^m+\hat\nabla_{\!r}J^{\,m}_j
-\hat\nabla_{\!j}J^{\,m}_r\right)\hat G_{mi}.
\endgathered\qquad
\mytag{4.16}
$$
As for the time derivatives $\Dot{\Hat T}\vphantom{T}^s_{jr}$,
$\Dot{\Hat T}\vphantom{T}^s_{ir}$, and $\Dot{\Hat T}\vphantom{T}^k_{ij}$,
we use the formula \mythetag{2.8} differentiating it with respect to $t$.
As a result, applying \mythetag{4.4}, we obtain
$$
\Dot{\Hat T}\vphantom{T}^k_{ij}=\sum^3_{r=1}\sum^3_{s=1}
\omega_{sij}\ g^{sr}\!
\left(\,\shave{\sum^3_{n=1}}\hat\nabla_{\!n}v^k\,R^n_r-
\shave{\sum^3_{n=1}}\shave{\sum^3_{\alpha=1}}
\shave{\sum^3_{\beta=1}}g_{rn}\ \omega^{n\alpha\beta}\ 
\hat\nabla_{\!\alpha}J^{\,k}_\beta\right)\!.
$$
The above huge formula can be simplified to the following one:
$$
\hskip -2em
\Dot{\Hat T}\vphantom{T}^k_{ij}=\hat\nabla_{\!j}J^{\,k}_i
-\hat\nabla_{\!i}J^{\,k}_j
+\sum^3_{n=1}\hat\nabla_{\!n}v^k\,\hat T^n_{ij}.
\mytag{4.17}
$$
From \mythetag{4.17} we easily derive the following two equalities:
$$
\align
&\hskip -2em
\sum^3_{s=1}\hat G_{is}\,\Dot{\Hat T}\vphantom{T}^s_{jr}=\sum^3_{m=1}
\hat G_{im}\left(\hat\nabla_{\!r}J^{\,m}_j-\hat\nabla_{\!j}J^{\,m}_r+
\shave{\sum^3_{n=1}}\hat\nabla_{\!n}v^m\,\hat T^n_{jr}\right)\!,
\mytag{4.18}\\
&\hskip -2em
\sum^3_{s=1}\hat G_{js}\,\Dot{\Hat T}\vphantom{T}^s_{ir}=\sum^3_{m=1}
\hat G_{jm}\left(\hat\nabla_{\!r}J^{\,m}_i-\hat\nabla_{\!i}J^{\,m}_r+
\shave{\sum^3_{n=1}}\hat\nabla_{\!n}v^m\,\hat T^n_{ir}\right)\!.
\mytag{4.19}
\endalign
$$
Now we apply \mythetag{4.16}, \mythetag{4.17}, \mythetag{4.18},
and \mythetag{4.19} to \mythetag{4.15}. As a result we obtain 
$$
\hskip -2em
\gathered
\Dot{\Hat\Gamma}\vphantom{\Gamma}^k_{ij}=-\frac{\hat\nabla_{\!i}
\hat\nabla_{\!j}\,v^k\,+\hat\nabla_{\!j}\hat\nabla_{\!i}\,v^k}{2}
-\hat\nabla_{\!i}J^{\,k}_j+\frac{1}{2}\sum^3_{n=1}\hat\nabla_{\!n}v^k
\,\hat T^n_{ij}\,+\\
+\sum^3_{m=1}\sum^3_{r=1}\frac{\widetilde G^{kr}}{2}
\left([\hat\nabla_{\!r},\,\hat\nabla_{\!i}]v^m
-\shave{\sum^3_{n=1}}\hat\nabla_{\!n}v^m\,\hat T^n_{ir}
\right)\hat G_{mj}\,+\\
+\sum^3_{m=1}\sum^3_{r=1}\frac{\widetilde G^{kr}}{2}
\left([\hat\nabla_{\!r},\,\hat\nabla_{\!j}]v^m
-\shave{\sum^3_{n=1}}\hat\nabla_{\!n}v^m\,\hat T^n_{jr}
\right)\hat G_{mi}.
\endgathered
\mytag{4.20}
$$
The commutators $[\hat\nabla_{\!r},\,\hat\nabla_{\!i}]$ and
$[\hat\nabla_{\!r},\,\hat\nabla_{\!j}]$ in the above formula
\mythetag{4.20} can be calculated on the base of the well-known
differential-geometric formula
$$
\hskip -2em
[\hat\nabla_{\!i},\,\hat\nabla_{\!j}]X^k=\sum^3_{n=1}\hat R^k_{nij}
\ X^n-\sum^3_{n=1}\hat T^n_{ij}\ \hat\nabla_{\!n}X^k.
\mytag{4.21}
$$
Here $X^k$ and $X^n$ stand for the components of an arbitrary vector field,
while $\hat R^k_{nij}$ are the components of the curvature tensor given by
the formula \mythetag{2.9}. Choosing $\bold X=\bold v$ in \mythetag{4.21}
and substituting \mythetag{4.21} into \mythetag{4.20}, we derive
$$
\hskip -2em
\gathered
\Dot{\Hat\Gamma}\vphantom{\Gamma}^k_{ij}=-\frac{\hat\nabla_{\!i}
\hat\nabla_{\!j}\,v^k+\hat\nabla_{\!j}\hat\nabla_{\!i}\,v^k}{2}
-\hat\nabla_{\!i}J^{\,k}_j+\frac{1}{2}\sum^3_{n=1}\hat\nabla_{\!n}v^k
\,\hat T^n_{ij}\,+\\
+\sum^3_{m=1}\sum^3_{r=1}\sum^3_{n=1}\left(\frac{\widetilde G^{kr}\,
\hat R^m_{nri}\,v^n\,\hat G_{mj}}{2}+\frac{\widetilde G^{kr}\,
\hat R^m_{nrj}\,v^n\,\hat G_{mi}}{2}\right)\!.
\endgathered
\mytag{4.22}
$$\par
    Now we are ready to calculate the time derivative of the curvature
tensor $\hat\bold R$ due to the time evolution of tensor fields $\hat
\bold G$ and $\bold R$ given by the equations \mythetag{4.4} and 
\mythetag{4.7}. Differentiating \mythetag{2.9} with respect to $t$,
we get
$$
\hskip -2em
\frac{\partial\hat R^k_{q\,ij}}{\partial t}=\hat\nabla_{\!i}
\Dot{\Hat\Gamma}\vphantom{\Gamma}^k_{\!j\,q}-\hat\nabla_{\!j}
\Dot{\Hat\Gamma}\vphantom{\Gamma}^k_{\!i\,q}+\sum^3_{m=1}
\hat T^m_{ij}\,\Dot{\Hat\Gamma}\vphantom{\Gamma}^k_{m\,q}.
\mytag{4.23}
$$
Before substituting \mythetag{4.22} into \mythetag{4.23} we transform it
using the identity \mythetag{4.21} again:
$$
\hat\nabla_{\!j}\hat\nabla_{\!i}\,v^k=\hat\nabla_{\!i}
\hat\nabla_{\!j}\,v^k-\sum^3_{n=1}\hat R^k_{nij}\,v^n
+\sum^3_{n=1}\hat T^n_{ij}\,\hat\nabla_{\!n}v^k.
$$
Due to this identity, the equality \mythetag{4.22} now looks like
$$
\gathered
\Dot{\Hat\Gamma}\vphantom{\Gamma}^k_{ij}=-\hat\nabla_{\!i}\hat\nabla_{\!j}
\,v^k-\hat\nabla_{\!i}J^{\,k}_j+\frac{1}{2}\sum^3_{n=1}\hat R^k_{nij}\,v^n
\,+\\
+\sum^3_{m=1}\sum^3_{r=1}\sum^3_{n=1}\left(\frac{\widetilde G^{kr}\,
\hat R^m_{nri}\,v^n\,\hat G_{mj}}{2}+\frac{\widetilde G^{kr}\,
\hat R^m_{nrj}\,v^n\,\hat G_{mi}}{2}\right)\!.
\endgathered
$$
Continuing our calculations, below we shall omit the terms containing the
components of the curvature tensor and their spatial derivatives denoting
them by dots:
$$
\hskip -2em
\Dot{\Hat\Gamma}\vphantom{\Gamma}^k_{ij}=-\hat\nabla_{\!i}\hat\nabla_{\!j}
\,v^k-\hat\nabla_{\!i}J^{\,k}_j+\,\dots\ .
\mytag{4.24}
$$
Then, substituting \mythetag{4.24} into \mythetag{4.23}, we derive:
$$
\hskip -2em
\gathered
\frac{\partial\hat R^k_{q\,ij}}{\partial t}=-\hat\nabla_{\!i}
\hat\nabla_{\!j}\hat\nabla_{\!q}\,v^k-\hat\nabla_{\!i}\hat\nabla_{\!j}
J^{\,k}_q+\hat\nabla_{\!j}\hat\nabla_{\!i}\hat\nabla_{\!q}\,v^k\,+\\
+\,\hat\nabla_{\!j}\hat\nabla_{\!i}J^{\,k}_q-\sum^3_{m=1}\hat T^m_{ij}
\,\hat\nabla_{\!m}\hat\nabla_{\!q}\,v^k-\sum^3_{m=1}\hat T^m_{ij}
\,\hat\nabla_{\!m}J^{\,k}_q+\,\dots\ .
\endgathered
\mytag{4.25}
$$
In order to simplify this formula let's write \mythetag{4.25} as follows:
$$
\hskip -2em
\aligned
\frac{\partial\hat R^k_{q\,ij}}{\partial t}=&-[\hat\nabla_{\!i},
\,\hat\nabla_{\!j}]\left(\hat\nabla_{\!q}\,v^k+J^{\,k}_q\right)-\\
&-\sum^3_{m=1}\hat T^m_{ij}\ \nabla_{\!m}\!\left(\hat\nabla_{\!q}
\,v^k+J^{\,k}_q\right)+\,\dots\ .
\endaligned
\mytag{4.26}
$$
The further transformation of \mythetag{4.26} is based on the other
well-known formula of the differential geometry, which is similar to \mythetag{4.21}:
$$
\hskip -2em
[\hat\nabla_{\!i},\,\hat\nabla_{\!j}]X^k_q=\sum^3_{n=1}\hat R^k_{nij}
\ X^n_q-\sum^3_{n=1}\hat R^n_{qij}\ X^k_n-\sum^3_{n=1}\hat T^n_{ij}\ 
\hat\nabla_{\!n}X^k_q.
\mytag{4.27}
$$
Here $X^k_q$ stand for the components of an arbitrary tensorial field
of the type $(1,1)$. Substituting $X^k_q=\hat\nabla_{\!q} \,v^k+J^{\,k}_q$
into \mythetag{4.27}, we see that \mythetag{4.26} can be written as
$$
\frac{\partial\hat R^k_{q\,ij}}{\partial t}=-\sum^3_{n=1}\hat R^k_{nij}
\left(\hat\nabla_{\!q}\,v^n+J^{\,n}_q\right)
+\sum^3_{n=1}\hat R^n_{qij}\left(\hat\nabla_{\!n}
\,v^k+J^{\,k}_n\right)+\,\dots\ .\quad
\mytag{4.28}
$$
Looking at \mythetag{4.28}, we see that two terms explicitly written in
the right hand side of this formula contain the components of the curvature
tensor $\hat\bold R$. \pagebreak According to the above our convention,
they should also be denoted by dots. This means that all terms in the right
hand side of \mythetag{4.28} would vanish provided the zero-curvature
condition \mythetag{2.16} is fulfilled.
\mytheorem{4.1} The time evolution of the tensor fields $\hat\bold G$ and
$\bold R$ given by the equations \mythetag{4.4} and \mythetag{4.7} preserves
the zero-curvature condition \mythetag{2.16}, i\.\,e\. the curvature tensor
$\hat\bold R$ is permanently equal to zero if \ $\hat\bold R=0$ at some
initial instant of time.
\endproclaim
Passing from $\hat{\boldsymbol\Gamma}$ back to the standard Euclidean
connection $\Gamma$ in the real space $\Bbb E$, and hence, from
$\hat\nabla$ back to $\nabla$, we can formulate the
theorem~\mythetheorem{4.1} as follows.
\mytheorem{4.2} The time evolution of the tensor fields $\hat\bold G$ and
$\bold R$ given by the equations \mythetag{1.1} and \mythetag{4.1} preserves
the zero-curvature condition \mythetag{2.16}, i\.\,e\. the curvature tensor
$\hat\bold R$ is permanently equal to zero if \ $\hat\bold R=0$ at some
initial instant of time.
\endproclaim
\head
5. Zero-divergency condition.
\endhead
     The zero-curvature condition \mythetag{2.16} is not the only condition
the tensor fields $\hat\bold G$ and $\bold R$ should obey. Another condition
was derived in \mycite{7} from the equality $\divr\boldsymbol\rho=0$. 
Therefore, we call it the {\it zero-divergency condition}:
$$
\hskip -2em
\sum^3_{p=1}\sum^3_{q=1}g^{p\,q}\,\nabla_{\!p}R^k_q
+\sum^3_{m=1}\sum^3_{p=1}\sum^3_{q=1}
g^{pq}\,R^m_q\,\hat Z^k_{pm}=0
\mytag{5.1}
$$
(see \thetag{4.8} in \mycite{7}). The following theorem shows that
\mythetag{5.1} is not an independent condition for $\hat\bold G$ and 
$\bold R$.
\mytheorem{5.1} The zero-divergency condition \mythetag{5.1} can be 
derived from the zero-curvature condition \mythetag{2.16}.
\endproclaim
     Formulating the theorem~\mythetheorem{5.1}, we assume that the 
tensor fields $\hat\bold G$ and $\bold R$ are given and that the 
following conditions are fulfilled:
\roster
\rosteritemwd=5pt
\item $\bold R$ define the torsion tensor $\hat\bold T$ according 
to the equality \mythetag{2.8};
\item $\hat\bold G$ and $\bold R$ define the non-symmetric connection
$\hat{\boldsymbol\Gamma}$ with the torsion $\hat\bold T$ according to 
the theorem~\mythetheorem{3.1};
\item the curvature tensor $\hat\bold R$ of the connection $\hat{\boldsymbol\Gamma}$ is equal to zero;
\item the tensor field $\hat\bold Z$ is determined as the difference
of the connection $\hat{\boldsymbol\Gamma}$ and the standard symmetric
Euclidean connection $\boldsymbol\Gamma$ according to the formula
\mythetag{2.3}.
\endroster
    From the above four conditions \therosteritem{1}--\therosteritem{4}
one easily derives the equality
$$
\hskip -2em
R^k_q=\sum^3_{n=1}\sum^3_{r=1}\sum^3_{s=1}
g_{qn}\ \omega^{nrs}\ \hat Z^k_{rs}.
\mytag{5.2}
$$
and the formula \mythetag{2.11} for the components of the curvature
tensor $\hat\bold R$. The zero-curvature condition $\hat\bold R=0$
then is equivalent to the symmetry condition
$$
\pagebreak
\hskip -2em
U^k_{sp\,r}=U^k_{s\,rp},
\mytag{5.3}
$$
where the quantities $U^k_{sp\,r}$ are given by the formula
$$
\hskip -2em
U^k_{sp\,r}=\nabla_{\!p}\hat Z^k_{r\,s}
+\sum^3_{m=1}\hat Z^m_{r\,s}\ \hat Z^k_{p\,m}.
\mytag{5.4}
$$
Due to the symmetry condition \mythetag{5.3} the following sum is obviously
equal to zero:
$$
\hskip -2em
\sum^3_{p=1}\sum^3_{r=1}\sum^3_{s=1}\omega^{p\,rs}\,U^k_{sp\,r}=0.
\mytag{5.5}
$$
Indeed, \ $U^k_{sp\,r}$ is symmetric, while $\omega^{p\,rs}$ is 
skew-symmetric with respect to $p$ and $r$. The next transformation
of the identity \mythetag{5.5} is also quite obvious:
$$
\hskip -2em
\sum^3_{q=1}\sum^3_{n=1}\sum^3_{p=1}\sum^3_{r=1}\sum^3_{s=1}
g^{p\,q}\,g_{qn}\,\omega^{n\,rs}\,U^k_{sp\,r}=0.
\mytag{5.6}
$$
Now it is sufficient to substitute \mythetag{5.4} into \mythetag{5.6}.
As a result we obtain
$$
\hskip -2em
\gathered
\sum^3_{p=1}\sum^3_{q=1}g^{p\,q}
\sum^3_{n=1}\sum^3_{r=1}\sum^3_{s=1}
\,g_{qn}\,\omega^{n\,rs}\,\nabla_{\!p}\hat Z^k_{r\,s}\,+\\
+\sum^3_{p=1}\sum^3_{q=1}g^{p\,q}\sum^3_{m=1}
\left(\,\shave{\sum^3_{n=1}\sum^3_{r=1}\sum^3_{s=1}}
\,g_{qn}\,\omega^{n\,rs}\,\hat Z^m_{r\,s}\right)\hat Z^k_{p\,m}=0.
\endgathered
\mytag{5.7}
$$
Applying \mythetag{5.2} to \mythetag{5.7}, we find that the equality
\mythetag{5.7} is equivalent to the zero-divergency condition
\mythetag{5.1}. Thus, the theorem~\mythetheorem{5.1} is proved.
\head
6. Conclusions.
\endhead
     As a main result we can formulate the following statement: the
elastic deformation tensor $\hat\bold G$ and the tensor of Burgers 
vector density $\bold R$ are two basic tensor fields describing 
completely the deformation state of a crystal. The time evolution
of $\hat\bold G$ and $\bold R$ is determined by the differential 
equations \mythetag{4.4} and \mythetag{4.7}\footnote{ \ Or by the 
equations \mythetag{4.1} and \mythetag{1.1}, which are equivalent to 
\mythetag{4.4} and \mythetag{4.7}.}. However, \mythetag{4.4} and
\mythetag{4.7} do not form the closed system of differential 
equations --- they are only the kinematic equations. They should be
completed with the dynamic equations relating $\bold J$ (the density 
of the Burgers vector flow) to $\hat\bold G$ and $\bold R$. 
Qualitatively, the process of a crystal deformation is expressed 
by the following diagram: \vadjust{\vskip 0pt\hbox to 0pt{\kern 80pt
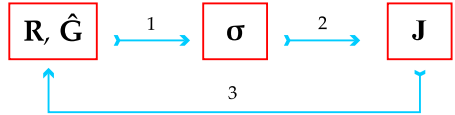
\includegraphics{diagram.eps}\hss}\vskip 70pt}\par
\noindent
The elastic deformation $\hat\bold G$ produces the stress $\boldsymbol
\sigma$, this phenomenon is expressed by the arrow 1 (Hooke's law or
its nonlinear generalization). The stress $\boldsymbol\sigma$ causes
the dislocations to move producing their flow $\bold J$, see the arrow 2
on the above diagram. And finally, the moving dislocations rearrange the
interatomic bonds causing $\hat\bold G$ and the stress $\boldsymbol\sigma$
to relax. This phenomenon is expressed by the arrow 3 and described by the
equations \mythetag{4.4} and \mythetag{4.7}. The plastic deformation tensor
$\check\bold G$ is not presented on the diagram. However, it can be
implicitly present in the arrows 1 and 2. The detailed quantitative description of the phenomena associated with these arrows is the subject
of the separate paper.\par


\Refs
\ref\myrefno{1}
\by Lyuksyutov~S.~F., Sharipov~R.~A.\paper Note on kinematics,
dynamics, and thermodynamics of plastic glassy media
\publ e-print \myhref{http://arXiv.org/abs/cond-mat/0304190/}
{cond-mat/0304190} in Electronic Archive \myEarXivlink
\endref
\ref\myrefno{2}\by Lyuksyutov~S.~F., Sharipov~R.~A.\paper
Separation of plastic deformations in polymers based on elements 
of general nonlinear theory
\publ e-print \myhref{http://arXiv.org/abs/cond-mat/0408433/}
{cond-mat/0408433} in Electronic Archive\linebreak\myEarXivlink
\endref
\ref\myrefno{3}\by Falk~M.~L., Langer~J.~S.\paper Dynamics of 
viscoplastic deformation in amorphous solids \jour Phys\. Rev\.~E
\vol 57\pages 7192--7205\yr 1998\moreref see also e-print
\myhref{http://arXiv.org/abs/cond-mat/9712114/}{cond-mat/9712114}
in Electronic Archive\linebreak\myEarXivlink
\endref
\ref\myrefno{4}\by Langer~J.~S., Lobkovsky~A.~E.\paper Dynamic ductile 
to brittle transition in a one-dimensional model of viscoplasticity
\jour Phys\. Rev\.~E\vol 58\yr 1998\pages 1568-1576
\endref
\ref\myrefno{5}\by Comer~J., Sharipov~R.~A.\paper A note on the 
kinematics of dislocations in crystals\publ e-print \myhref{http://arXiv.org/abs/math-ph/0410006/}{math-}\linebreak
\myhref{http://arXiv.org/abs/math-ph/0410006/}{ph/0410006}
in Electronic Archive \myEarXivlink
\endref
\ref\myrefno{6}
\by Sharipov~R.~A.\paper Gauge or not gauge\,?\publ 
e-print \myhref{http://arXiv.org/abs/cond-mat/0410552/}{cond-mat/0410552}
in Electronic Archive\linebreak\myEarXivlink
\endref
\ref\myrefno{7}
\by Sharipov~R.~A.\paper Burgers space versus real space in the nonlinear
theory of dislocations\publ e-print
\myhref{http://arXiv.org/abs/cond-mat/0411148/}{cond-mat/0411148}
in Electronic Archive \myEarXivlink
\endref
\ref\myrefno{8}
\by Sharipov~R.~A.\book Quick introduction to tensor analysis
\publ free on-line textbook in Electronic Archive \myEarXivlink;
see \myhref{http://arXiv.org/abs/math.HO/0403252}{math.HO/0403252}
and \myhref{http://www.geocities.com/r-sharipov/r4-b6.htm}
{r-sharipov/r4-b6.htm} in \myGeoCities
\endref
\ref\myrefno{9}
\by Sharipov~R.~A.\book Course of differential geometry\publ
Bashkir State University\publaddr Ufa\yr 1996\moreref see also
\myhref{http://arXiv.org/abs/math.HO/0412421}{math.HO/0412421}
in Electronic Archive \myEarXivlink\ and 
\myhref{http://www.geocities.com/r-sharipov/r4-b3.htm}
{r-sharipov/r4-b3.htm} in \myhref{http://www.geocities.com}{Geo-}
\myhref{http://www.geocities.com}{Cities}
\endref
\ref\myrefno{10}
\by Unzicker~A.\paper Teleparallel space-time with defects yields 
geometrization of electrodynamics with quantized charges\publ e-print
\myhref{http://arXiv.org/abs/gr-qc/9612061/}{gr-qc/9612061}
in Electronic Archive \myEarXivlink
\endref
\ref\myrefno{11}
\by Kondo~K.\book RAAG Memoirs of the unifying study of the basic
problems in physics and engineering science by means of geometry
\publ Volume 1, Gakujutsu Bunken Fukyu-Kay\publaddr Tokio\yr 1952
\moreref\publ Volume 2\yr 1955
\endref
\ref\myrefno{12}
\by Bilby~B.~A., Bullough~R., Smith~E.\paper Continuous distributions
of dislocations: a new application of the methods of non-Riemannian
geometry\jour Proc\. Royal Soc\. London\vol 231\,A\yr 1955
\pages 263--273
\endref
\ref\myrefno{13}
\by Sharipov~R.~A.\book Dynamical systems admitting the normal shift
\publ thesis for the degree of Doctor of Sciences in Russia\yr 2000
\moreref see \myhref{http://arXiv.org/abs/math/0002202/}{math.DG/0002202}
in Electronic Archive \myEarXivlink
\endref
\ref\myrefno{14}
\by Sharipov~R.~A.\paper On the point transformations for the equation
$y''=P+ 3\,Q\,y'+3\,R\,y'{}^2+S\,y'{}^3$\publ e-print
\myhref{http://arXiv.org/abs/solv-int/9706003/}{solv-int/9706003}
in Electronic Archive \myEarXivlink\moreref see also the short 
Russian version in Vestnik BashGU \vol 1\issue 1\yr 1998
\endref
\endRefs
\enddocument
\end